\documentclass[journal]{IEEEtran}
\usepackage{ifpdf}

\usepackage{cite}

\ifCLASSINFOpdf
  \usepackage[pdftex]{graphicx}
\else
  \usepackage[dvips]{graphicx}
\fi
\usepackage{amsmath}
\usepackage{array}

\ifCLASSOPTIONcompsoc
  \usepackage[caption=false,font=normalsize,labelfont=sf,textfont=sf]{subfig}
\else
  \usepackage[caption=false,font=footnotesize]{subfig}
\fi

\usepackage{booktabs}
\usepackage{color}
\usepackage{soul}

\begin{document}
%
\title{A CNN Accelerator on FPGA Using Depthwise Separable Convolution}
%
%
%

\author{Lin~Bai,~\IEEEmembership{Student~Member,~IEEE,}
        Yiming~Zhao,
        and~Xinming~Huang,~\IEEEmembership{Senior~Member,~IEEE}
    \thanks{This work was supported in part by the U.S. NSF under Grant 1626236 and by The MathWorks.}
	\thanks{The authors are with the Department of Electrical and Computer Engineering, Worcester Polytechnic Institute, MA 01609, USA. The corresponding aurthor is X. Huang (e-mail:xhuang@wpi.edu)}
}

\maketitle

\begin{abstract}
Convolutional neural networks (CNNs) have been widely deployed in the fields of computer vision and pattern recognition because of their high accuracy.
However, large convolution operations are computing intensive and often require a powerful computing platform such as a Graphics Processing Unit (GPU). This makes it 
difficult to apply CNNs to portable devices. The state-of-the-art CNNs, such as MobileNetV2 and Xception, adopt depthwise separable convolution to replace the standard convolution for embedded platforms, which significantly reduces operations and parameters with only limited loss in accuracy. This highly structured model is very suitable for Field-Programmable Gate Array (FPGA) implementation. 
In this paper, a scalable high performance depthwise separable 
convolution optimized CNN accelerator is proposed. The accelerator can be fit into an FPGA of different sizes, provided the balancing between hardware resources and processing speed. \textcolor{black}{As an example, MobileNetV2 is implemented on Arria 10 SoC FPGA,  and the results show this accelerator can classify each picture from ImageNet in 3.75ms, which is about 266.6 frames per second. The FPGA design achieves 20x speedup if compared to CPU}.
\end{abstract}


%
\IEEEpeerreviewmaketitle


\section{Introduction}\label{sec:intro}
%
%
%
%
\IEEEPARstart{N}{owadays}, convolutional neural networks (CNNs) have become the center of interest, due to their superior performance in tasks ranging from image 
classification, semantic segmentation, to object detection
and 
tracking. This technique has also been widely used in the industry, such as autonomous driving, video surveillance, speech recognition, etc.
\par
CNN is a computing intensive model. It consumes huge amounts of computing power during training and deployment. In practice, 
Graphics Processing Units (GPUs) are often selected as the platform. However, GPU's natural of high power consumption limits its application in 
embedded scenario such as portable devices and wearable systems. Therefore, Field-Programmable Gate Arrays (FPGAs) and Application-Specific Integrated Circuits (ASICs), as the replacement 
of GPUs, are adopted in neural network applications~\cite{chen2017eyeriss, chen2014dadiannao, du2015shidiannao, xiao2017exploring, venieris2017fpgaconvnet, li2016high, ma2017optimizing, qiu2016going, Tapiador2016, Aimar2017, Jiang2018, Zhao2018}. 
More specifically, increasing research attention is focused on FPGA-based CNN accelerator due to the possibility of trade-off between power consumption and reconfigurability.
\par
To further lighten the computing burden of standard convolution, depthwise separable convolution is 
proposed in~\cite{sifre2014rigid}. This has been applied in MobileNetV1~\cite{howard2017mobilenets} and later MobileNetV2 \cite{sandler2018inverted}, and thus achieved 
comparable results with much less multiply-accumulation operations and parameters.
\par
Almost all the existed FPGA-based CNN implementation works were to explore memory bandwidth and computing parallelism limitations. To conquer the limitation of memory bandwidth, \cite{chen2014dadiannao} and \cite{du2015shidiannao} stored the parameters in on-chip memory. 
However, as CNN goes deeper, parameters required by convolution increase sharply, which makes the on-chip memory solution inefficient. Other works like
\cite{xiao2017exploring}\cite{venieris2017fpgaconvnet}\cite{li2016high} alleviated the pressure on off-chip memory through limiting the parameters precision of the neural networks, as lower 
numerical precision were proved to be sufficient for CNN\cite{courbariaux2014low}\cite{gupta2015deep}. In \cite{ma2017optimizing}\cite{qiu2016going}, computing engine was optimized for highly parallelism in computation. \cite{li2016high} proposed a pipeline based solution for CNN for high throughput.
\textcolor{black}{\cite{Tapiador2016} made a comprehensive evaluation and comparison of Altera and Xilinx OpenCL frameworks for CNN.} 
\textcolor{black}{\cite{Aimar2017} explored the sparsity-based optimizations, which could achieve up to 3x higher core energy efficiency and raise the device-level energy efficiency by around 70\% through data compression.}
\textcolor{black}{Both \cite{Jiang2018} and \cite{Zhao2018} implemented separable depthwise convolution with the example MobileNetV1, and achieved processing speed at 7.85ms per image and 231.7 frames per second (fps) respectively.}
\par
The key contributions of this work are:
\par
(1) A high performance CNN hardware accelerator framework is proposed where all layers are processed in a computing unit named matrix multiplication engine.
\par
(2) The utilization of hierarchical memory structure and ping-pong on-chip buffer reduces the bandwidth limitation of off-chip memory.
\par
(3) A methodology for scalable design is proposed, so that this framework can be implemented in various FPGAs, through 
balancing the on-chip resources and performance.
\par
(4) By applying the proposed framework and methods, the state-of-the-art CNN, MobileNetV2~\cite{sandler2018inverted}, for the first time, is implemented on Arria 10 SoC FPGA. The results show 266.6 frames per second and 170.6 Giga Operations Per Second (GOPS) at system clock frequency of 133MHz. This represents a 20x speedup comparing to that on CPU~\cite{sandler2018inverted}.
\par
This paper is organized as follows. Section~\ref{sec:dsp_conv} provides fundamental knowledge of depthwise separable convolution, followed by
one of its application, MobilNetV2. Section~\ref{sec:system_design} describes the architecture of the accelerator, including the matrix multiplication engine, and 
on-chip buffer organization. System implementation and its results are discussed in Section~\ref{sec:system_impl}.
The conclusion is given in Section~\ref{sec:conclusion}.



\section{Depthwise Separable Convolution}\label{sec:dsp_conv}
\begin{figure}[htbp]
	\centering
	\subfloat[standard convolution]{\includegraphics[width=2.99in]{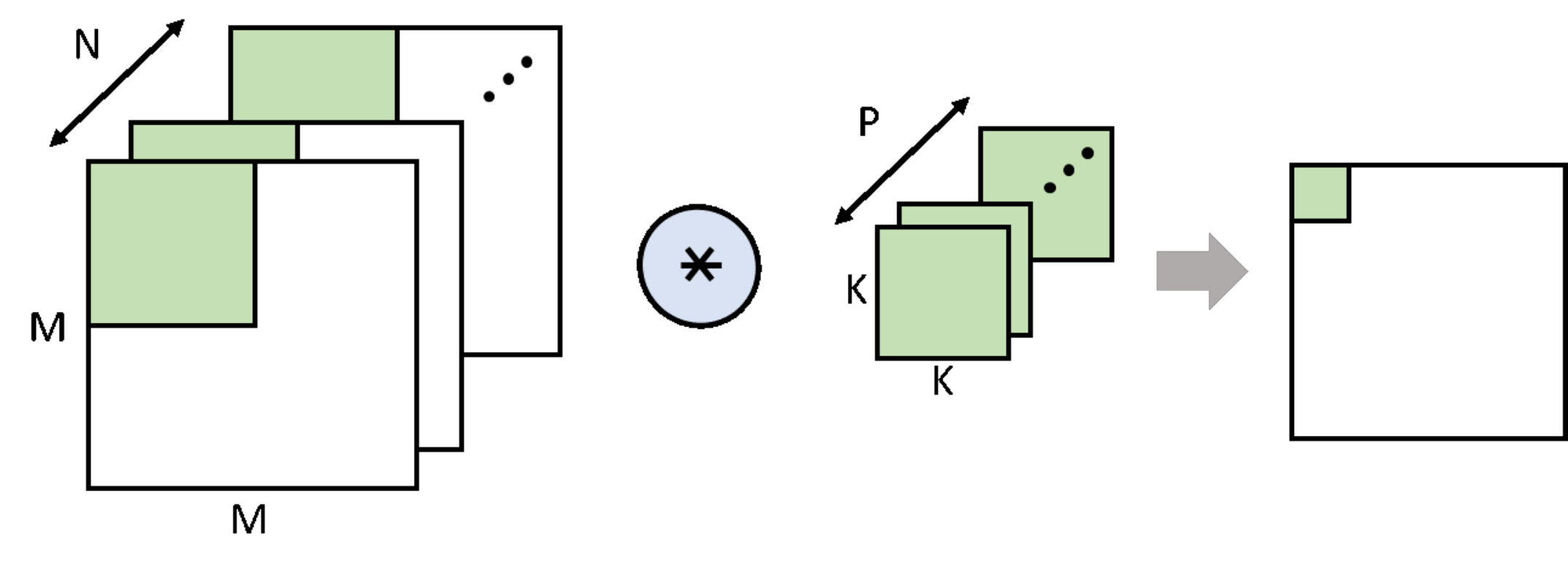}
		\label{fig:std_conv}}
	\hfil
	\subfloat[depthwise convolution]{\includegraphics[width=2.99in]{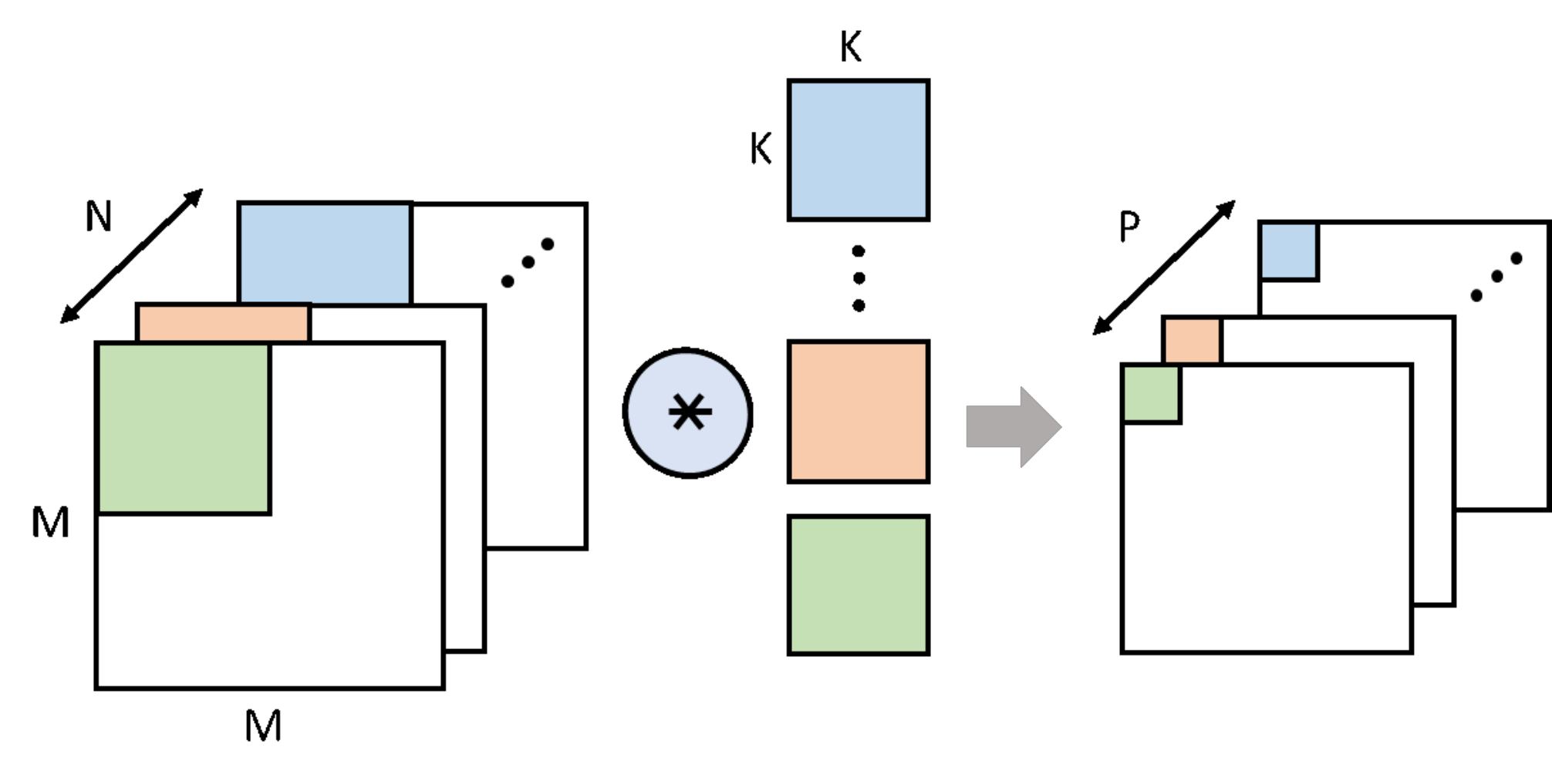}
		\label{fig:depth_conv}}
	\hfil
	\subfloat[pointwise convolution]{\includegraphics[width=2.99in]{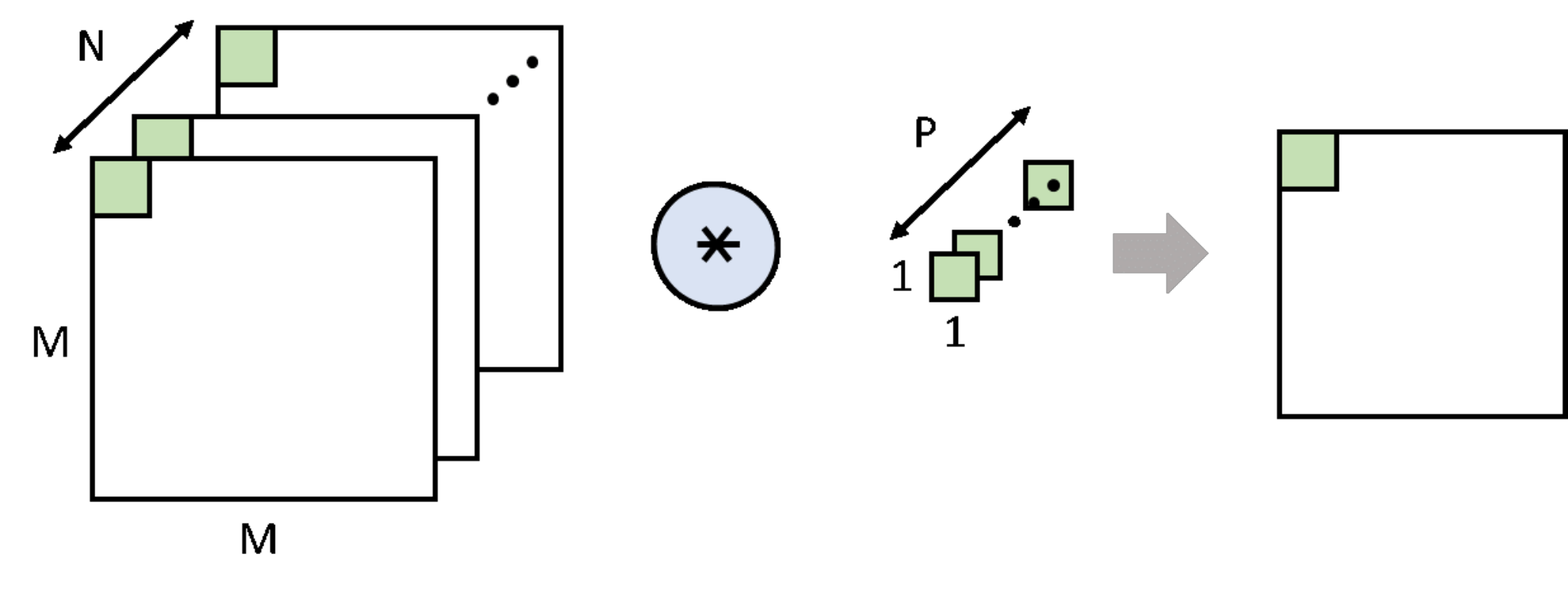}
		\label{fig:point_conv}}
	\caption{Comparison of different convolution types}
	\label{fig:convolution}
\end{figure}
Depthwise separable convolution was first introduced in \cite{chollet2016xception}. As one kind of the factorized convolutions, depthwise separable convolution factorizes the standard convolution
into a depthwise convolution plus a pointwise convolution. Fig.~\ref{fig:convolution} demonstrates how the standard convolution (SC), depthwise convolution (DWC) and pointwise convolution (PWC) work. In standard convolution, each input channel has to do a convolution with one specific kernel, and then the result is the sum of the convolution results from all channels.
While in depthwise separable convolution case, depthwise convolution is the first step, performing the convolution for each input channel individually. The next step is to do convolution in pointwise, which is actually a standard convolution with kernel size $1\times1$. Comparing to standard convolution, using depthwise separable convolution considerably reduces the number of mathematical operations and the number of parameters.
\par
As it is shown in Fig.~\ref{fig:convolution}, considering the input feature map with size $M\times M\times N$ and kernel size $K\times K\times N\times P$, in case of stride length of 1, the number of weights needed for standard convolution is \cite{howard2017mobilenets}
\begin{equation}
	W_{SC} = K\times K\times N\times P
\end{equation}
and the corresponding number of operations is
\begin{equation}
	O_{SC} = M\times M\times K\times K\times N\times P
\end{equation}
In case of depthwise separable convolution, the total number of weights is
\begin{equation}
	W_{DSC} = K\times K\times N + N\times P
\end{equation}
and the total number of operations is
\begin{equation}
	O_{DSC} = M\times M\times K\times K\times N + M\times M\times N\times P
\end{equation}

Thus, the reduction factors on weights and operation are calculated in (\ref{eq:factor_w}-\ref{eq:factor_o}):
\begin{equation}
	\label{eq:factor_w}
	F_W = \frac{W_{DSC}}{W_{SC}}= \frac{1}{P}+\frac{1}{K^2}
\end{equation}
\begin{equation}
	\label{eq:factor_o}
	F_O = \frac{O_{DSC}}{O_{SC}}= \frac{1}{P}+\frac{1}{K^2}
\end{equation}

\par
\begin{figure}[htbp]
	\centering
	\subfloat[stride = 1]{\includegraphics[width=1.2in]{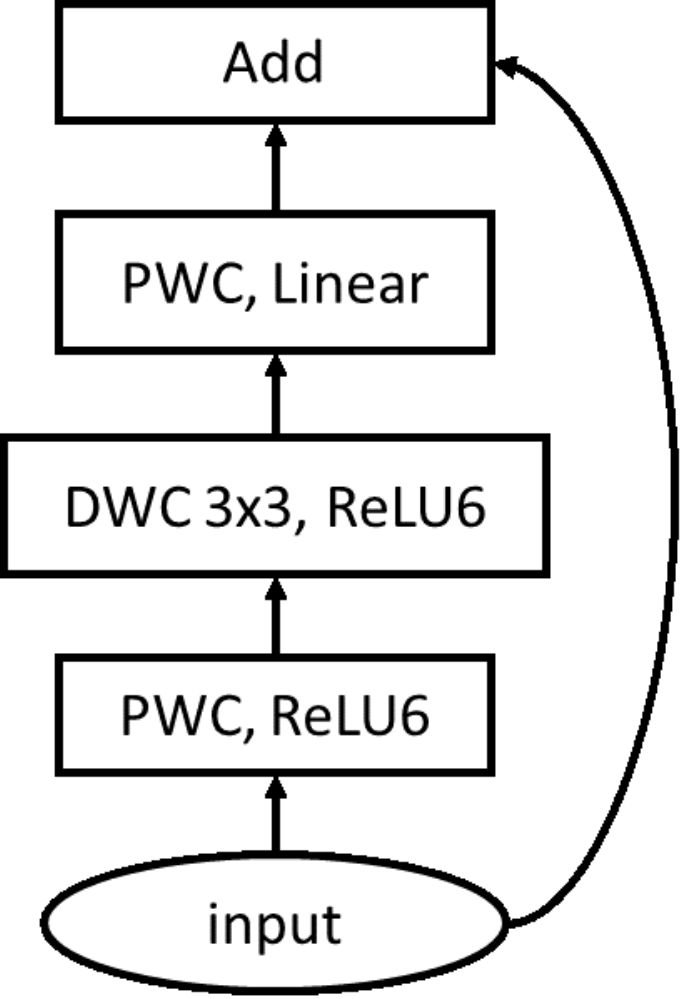}
	\label{fig:bottleneck_1}}
	\hfil
	\subfloat[stride = 2]{\includegraphics[width=1.0in]{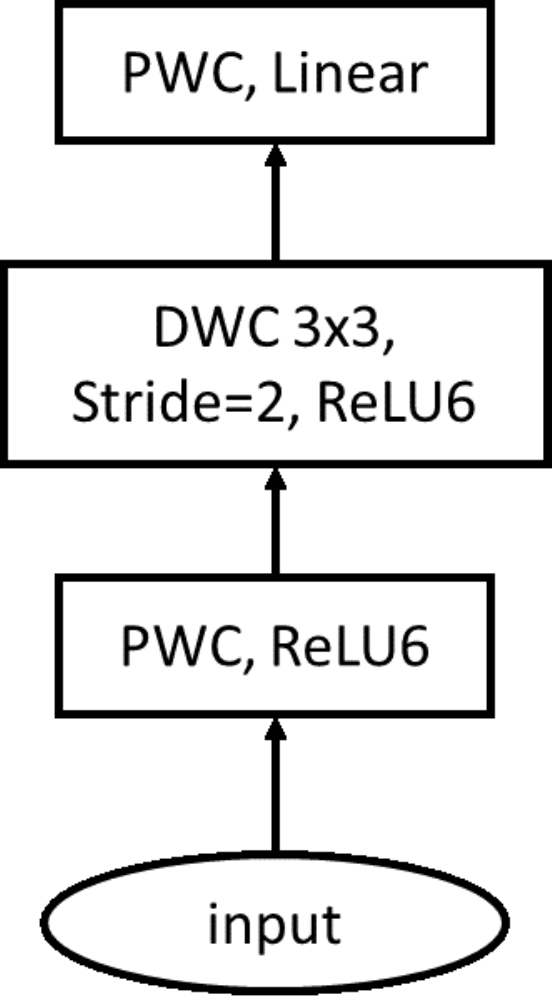}
	\label{fig:bottleneck_2}}
	\caption{Bottleneck operations in different strides}
	\label{fig:bottleneck}
\end{figure}

One of the typical application of depthwise separable convolution is MobileNetV2, the successor of MobileNetV1 \cite{howard2017mobilenets}. Comparing to its first version, the newly proposed MobileNetV2 further decreased the number of weights by shrinking the output channels in some layers. It also improves its performance through importing one more pointwise convolution layer before the depthwise separable convolution. The new operation is called bottleneck (Fig.~\ref{fig:bottleneck}).
\par
The network structure of MobileNetV2 is illustrated in Table~\ref{tab:mobile_struc}.
\begin{table}[htbp]
\centering
\caption{MobileNetV2 Structure~\cite{sandler2018inverted},  
	where each line represents a sequence
	of 1 or more identical (except stride) layers. All
	depthwise convolutions use 3x3 kernels.}
\label{tab:mobile_struc}
\begin{tabular}{c|c|c|c|c|c}
	\toprule[2pt]
	      &          & extend & output & repeat &  \\ 
	input & operator & factor & channel & time & stride \\ 
	\midrule[1pt]
	224x224x3 & standard conv. & - & 32 & 1 & 2\\ 
	112x112x3 & bottleneck & 1 & 16 & 1 & 1\\ 
	112x112x16 & bottleneck & 6 & 24 & 2 & 2\\ 
	56x56x24 & bottleneck & 6 & 32 & 3 & 2\\ 
	28x28x32 & bottleneck & 6 & 64 & 4 & 2\\ 
	14x14x64 & bottleneck & 6 & 96 & 3 & 1\\
	14x14x96 & bottleneck & 6 & 160 & 3 & 2\\ 
	7x7x160 & bottleneck & 6 & 320 & 1 & 1\\ 
	7x7x320 & pointwise conv. & - & 1280 & 1 & 1\\ 
	7x7x1280 & avgpool 7x7 & - & - & 1 & -\\ 
	1x1x1280 &pointwise conv. & - & 1000 & - & \\ 
	\bottomrule[2pt]
\end{tabular}
\end{table}

\section{System Design}\label{sec:system_design}
\subsection{Architecture Overview}
\begin{figure}[htbp]
	\centering
	\includegraphics[width=2.9in]{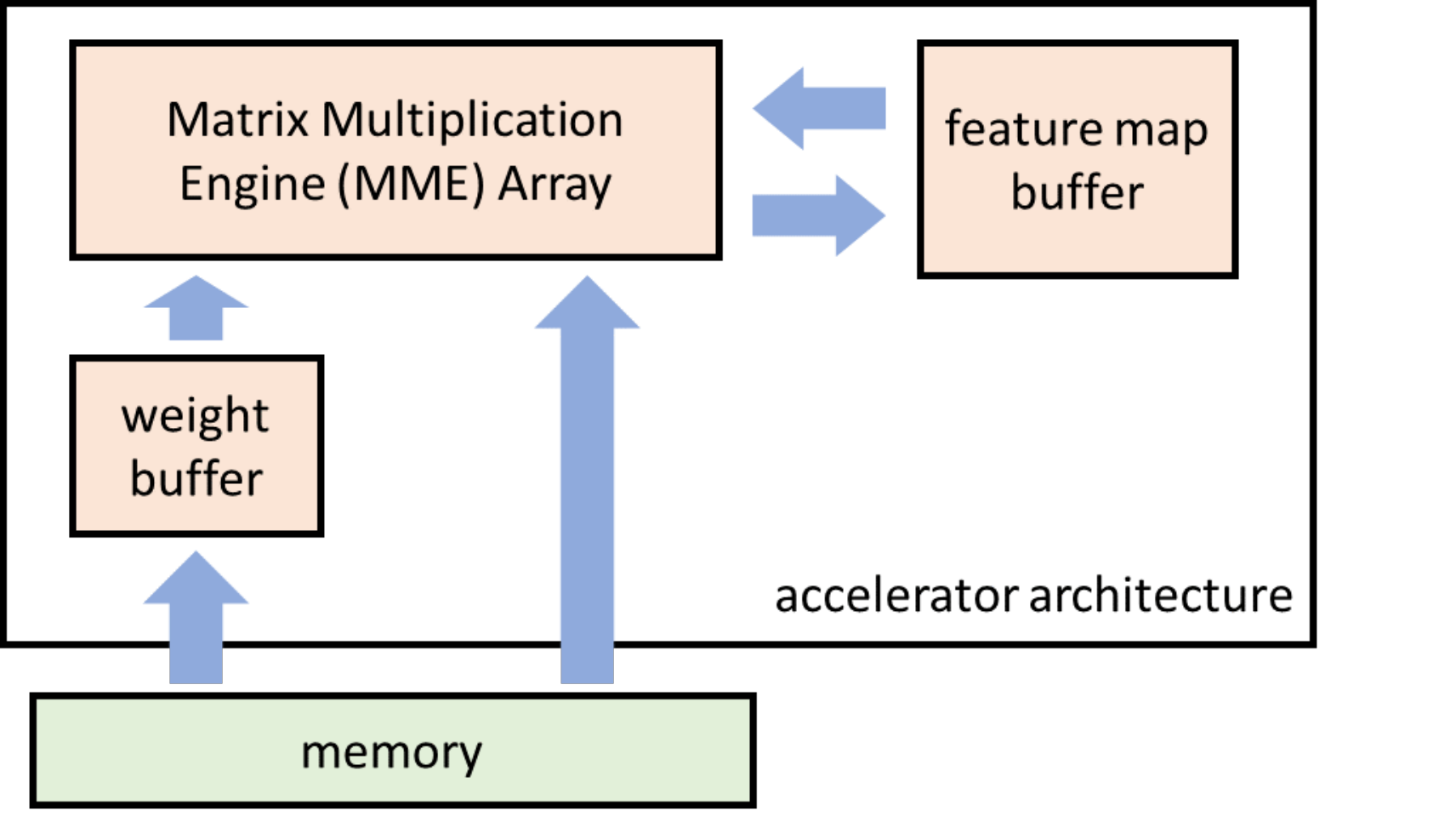}
	\caption{Block diagram of accelerator system}
	\label{fig:accel_arch}
\end{figure}

The block diagram in Fig.~\ref{fig:accel_arch} gives an overview of this accelerator. 
The proposed matrix multiplication engine (MME) array in this paper is responsible for all the CNN operations, including convolution, normalization, ReLU and pooling. All the parameters and input images are stored on off-chip memory. A ping-pong weight buffer is placed between MME array and memory to maximize the bandwidth. Biases are loaded to the registers in MME array. Feature map buffer stores all
the intermediate feature maps to avoid the latency brought by off-chip memory read and write. The accelerator is controlled by a general finite state machine (FSM).
\subsection{Matrix Multiplication Engine}
\begin{figure}[htbp]
	\centering
	\includegraphics[width=3.2in]{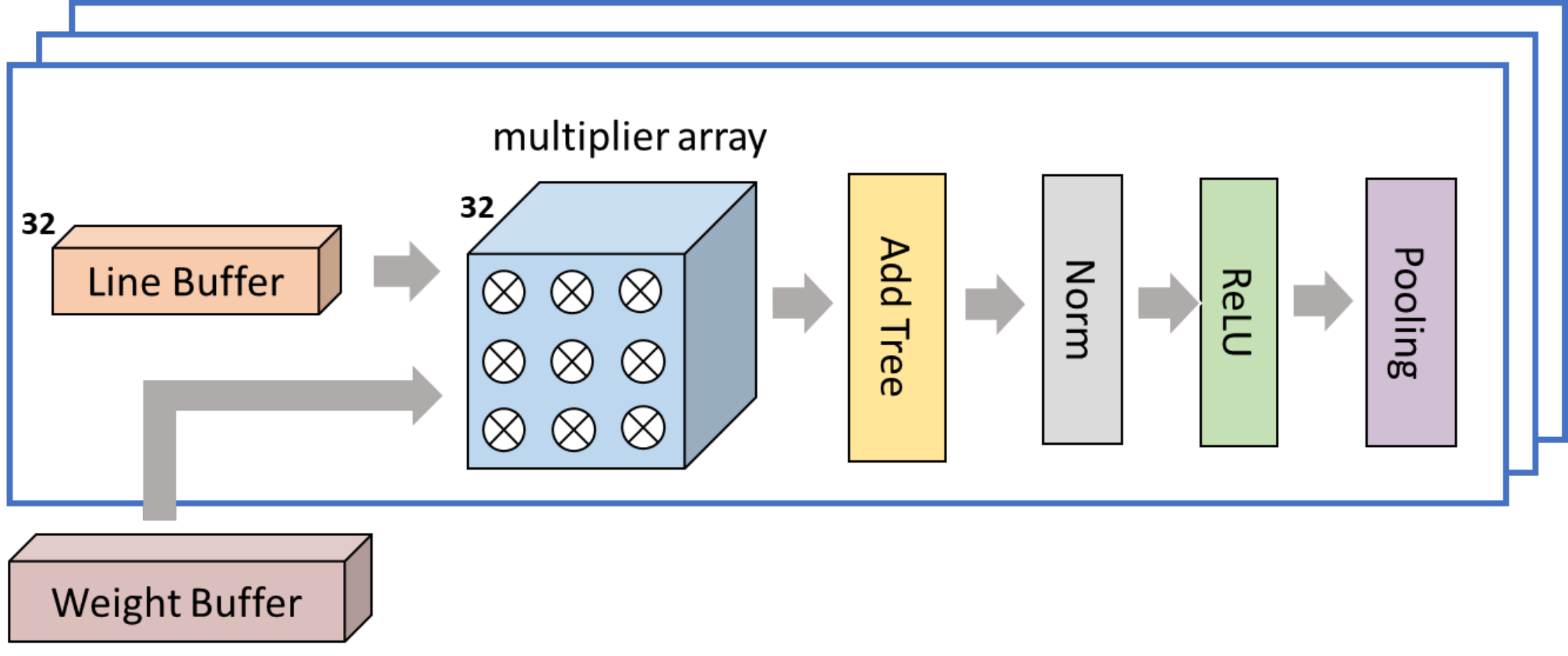}
	\caption{Block diagram of an MME}
	\label{fig:mme}
\end{figure}

In this paper, each MME consists of 32 slices line buffer,  32 slices $3\times 3$ multiplier array, 1 adder tree, 1 normalization (Norm) block, 1 ReLU block and 1 pooling block (Fig.~\ref{fig:mme}). In each convolution, MME loads the feature maps and 
weights to line buffers. After multiplication in multiplier array, adder tree sums the products according to the selected convolution type. The following operations are optional normalization, ReLU and pooling. 
\subsubsection{Line Buffer}
\begin{figure}[htbp]
	\centering
	\includegraphics[width=2.3in]{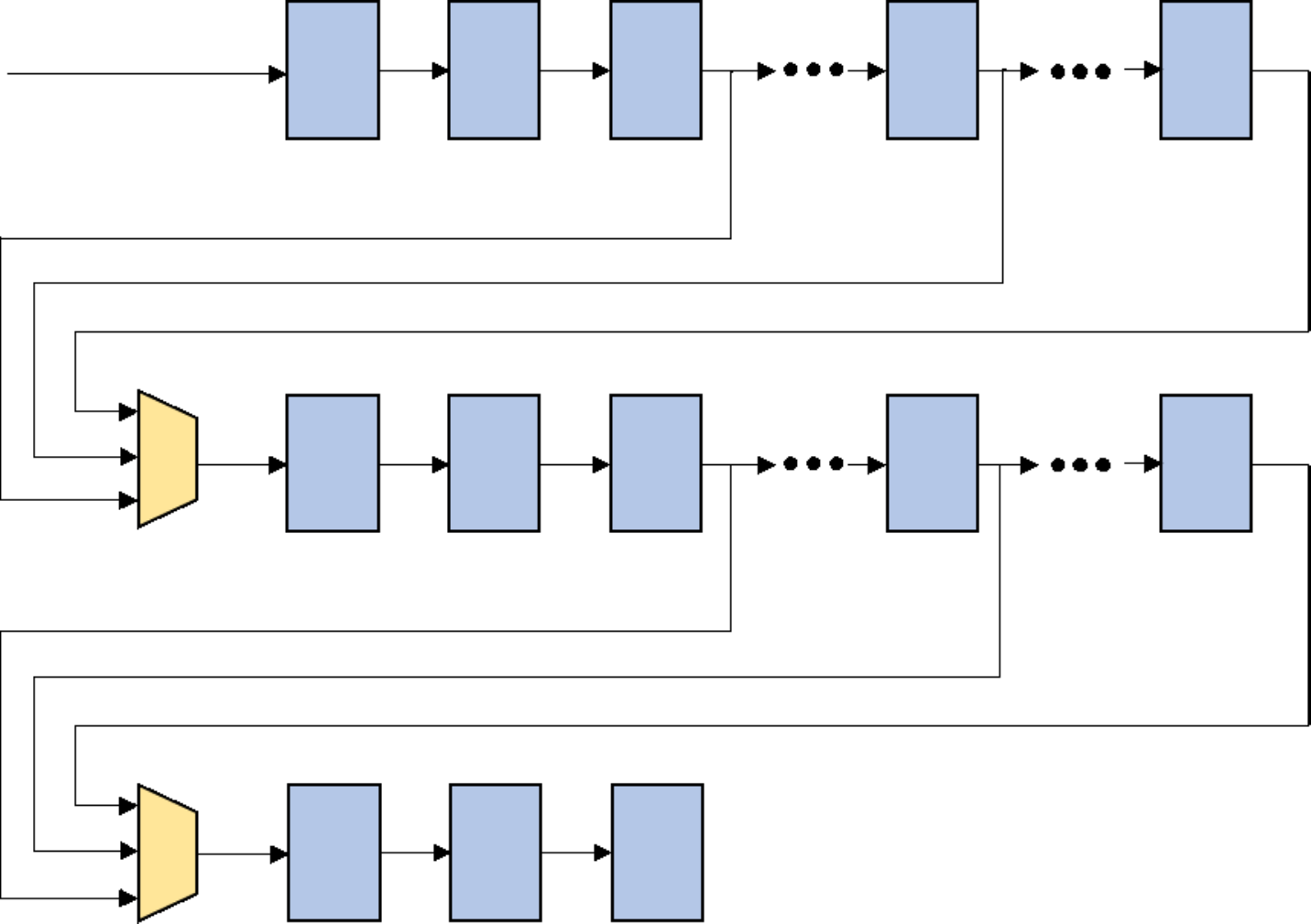}
	\caption{Line buffer in MME}
	\label{fig:line_buf}
\end{figure}

The working length of line buffer can be selected by control FSM to fit different input sizes, as it is illustrated by Fig.~\ref{fig:line_buf}. The implementation length is $(K-1)\times M+K$.
\subsubsection{Adder Tree}
\begin{figure}[htbp]
	\centering
	\subfloat[depthwise sum]{\includegraphics[width=1.0in]{./add_tree_1}
		\label{fig:add_tree_1}}
	\hfil
	\subfloat[pointwise sum]{\includegraphics[width=1.0in]{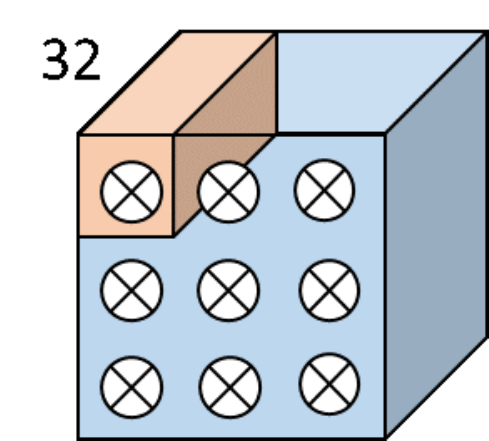}
		\label{fig:add_tree_2}}
	\caption{Adder tree modes for different convolution}
	\label{fig:add_tree_mode}
\end{figure}

\begin{figure}[!htb]
	\centering
	\includegraphics[width=3in]{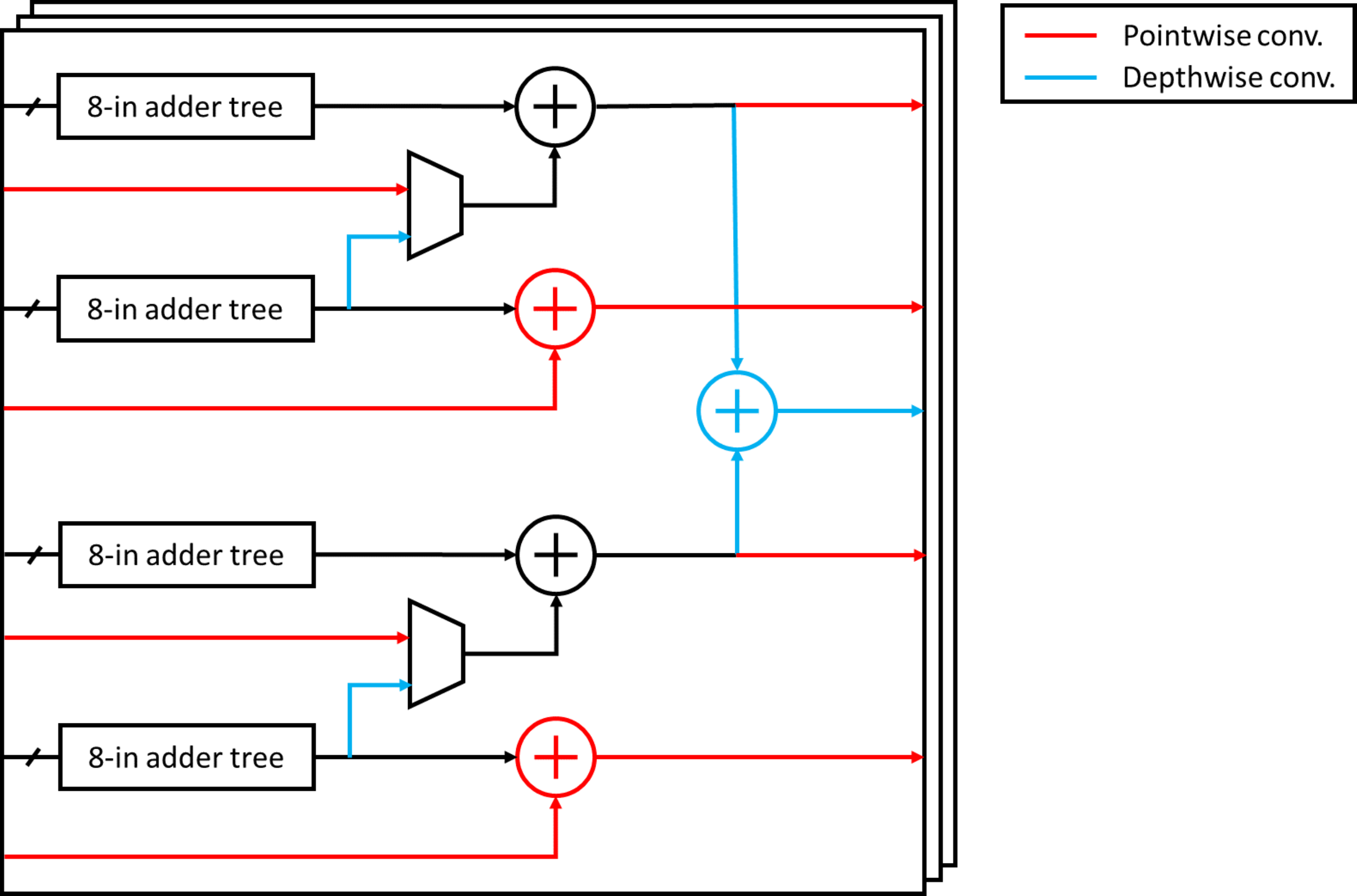}
	\caption{Block diagram of adder tree}
	\label{fig:add_tree}
\end{figure}

Adder tree is configurable to do the summing operation in depthwise or pointwise (Fig.~\ref{fig:add_tree_mode}). In Fig.~\ref{fig:add_tree}, black lines or blocks are shared by both types of convolution. Blue part is used 
when doing depthwise convolution. While red part works if pointwise convolution is selected. All the biases all added in this stage.
\par
\subsubsection{Standard Convolution}
To avoid losing too much information, standard convolution is adopted to do the first layer convolution. Therefore, this accelerator is adapted to be able to do the standard convolution with input feature map channel is 3.
For vision applications, the channel number of input feature map is always 3.
\subsubsection{Depthwise Convolution}
\begin{figure}[htbp]
	\centering
	\includegraphics[width=2.48in]{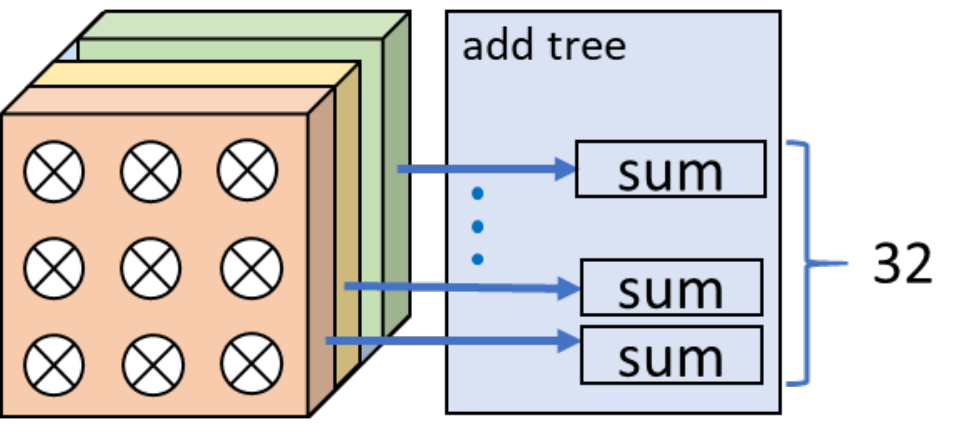}
	\caption{Depthwise convolution in MME}
	\label{fig:mme_depth_conv}
\end{figure}

Depthwise convolution performs convolution for each feature map separately. As shown in Fig.~\ref{fig:mme_depth_conv}, adder tree is configured to sum up the products from each slice of multiplier array in parallel. For one MME, the output channel number is 32.
\subsubsection{Pointwise Convolution}
\begin{figure}[htbp]
	\centering
	\includegraphics[width=2.5in]{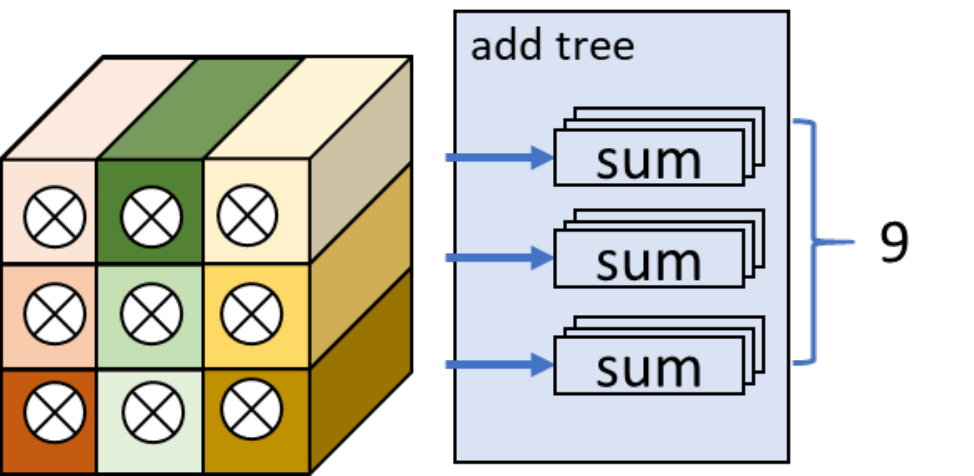}
	\caption{Pointwise convolution in MME}
	\label{fig:mme_point_conv}
\end{figure}

\begin{figure}[htbp]
	\centering
	\includegraphics[width=2.8in]{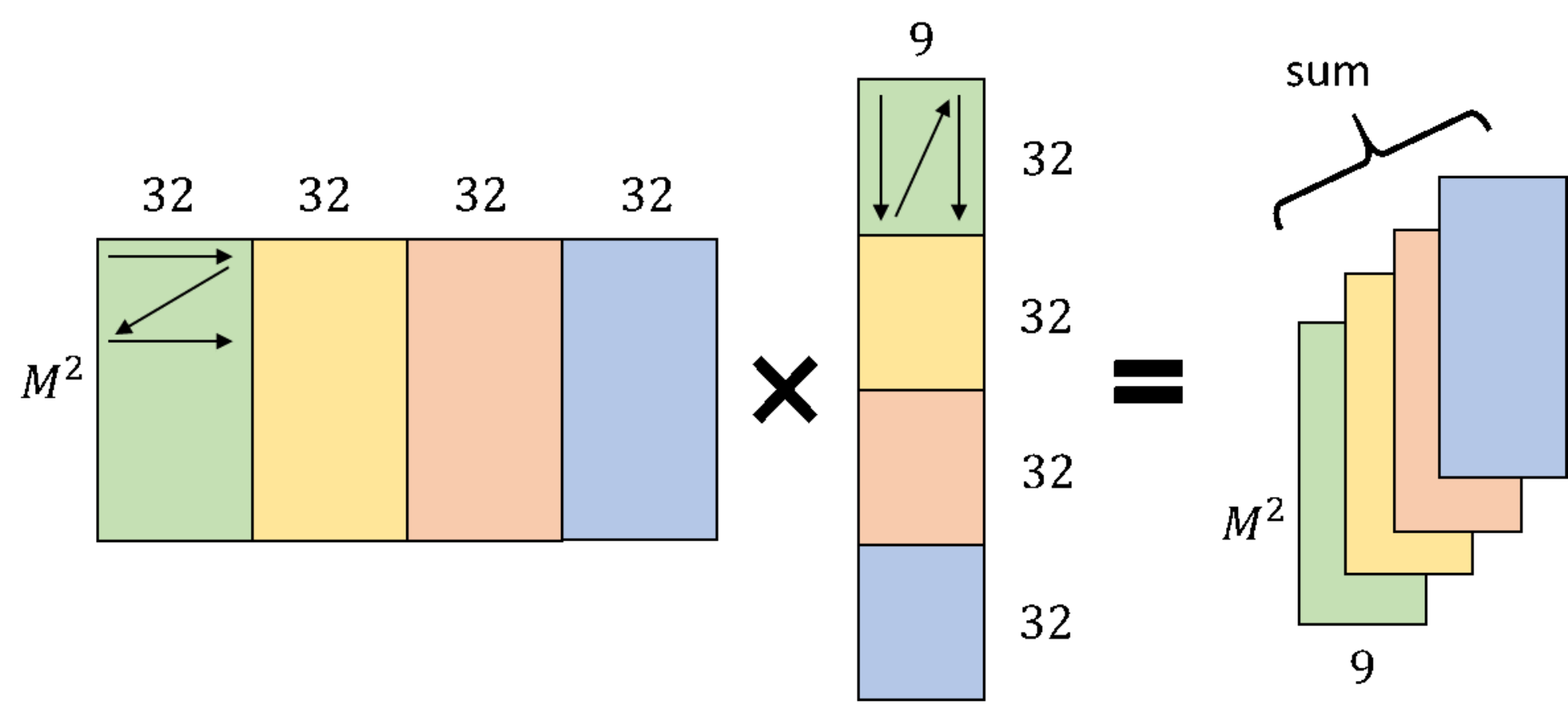}
	\caption{Divide-conquer for large matrix multiplication}
	\label{fig:div_conq}
\end{figure}

Pointwise convolution is actually standard convolution with kernel size $1\times 1$ (Fig.~\ref{fig:mme_point_conv}). To fully take advantage of all the multipliers in MME, the input feature map is divided into several $M\times M\times 32$ sub-matrices, and these sub-matrices are shifted into line buffers one after another. This idea comes from divide and conquer algorithm in large matrix multiplication illustrated in Fig.~\ref{fig:div_conq}, which consists in dividing large matrix into several small matrices and sum the results up after doing small matrix multiplication. For one MME, it is able to do $M^2\times 32$ and $32\times 9$ multiplication at once. The adder tree sums up the 32 products in each cell as revealed by Fig.~\ref{fig:mme_point_conv}. Thus the output channel number is 9.
\subsubsection{Normalization}
After training, parameters of batch normalization are fixed\cite{ioffe2015batch}. Thus the complex normalization is downgraded into multiplication and add operation.
\subsubsection{Pooling}
Average pooling and max pooling are treated differently. As pixels of a feature map channel are output one by one, average pooling could be easily calculated by adding one more multiply-accumulate stage 
by a factor of $1/S$, where $S$ is average pooling size. On the other hand, max pooling needs one more comparison stage.
\subsubsection{ReLU}
Same as the pooling layer, a ReLU stage is added after the normalization stage. Three options: no ReLU, standard ReLU and ReLU6 are selectable.

\subsection{Memory Organization}
\begin{figure}[htbp]
	\centering
	\subfloat[buffer 1 outputs, buffer 2 loads ]{\includegraphics[width=1.25in]{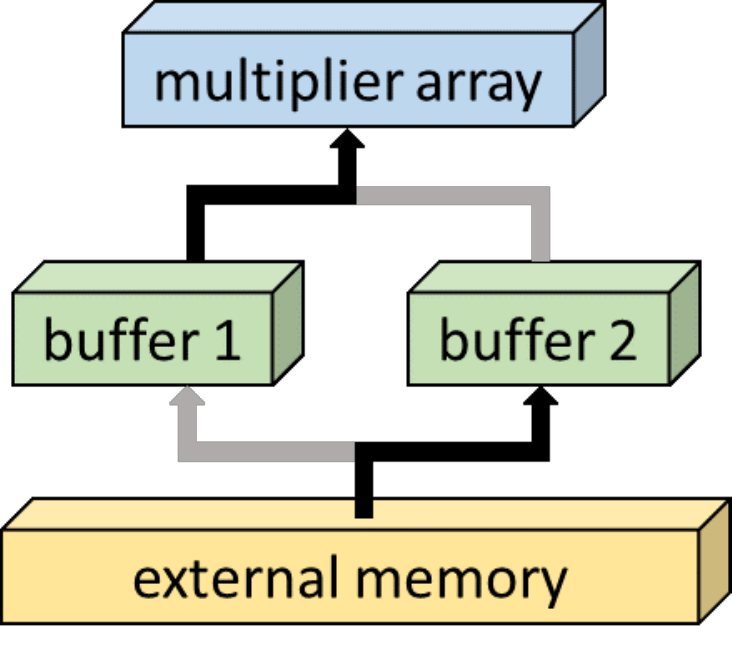}
		\label{fig:double_buffer_1}}
	\hfil
	\subfloat[buffer 2 outputs, buffer 1 loads]{\includegraphics[width=1.25in]{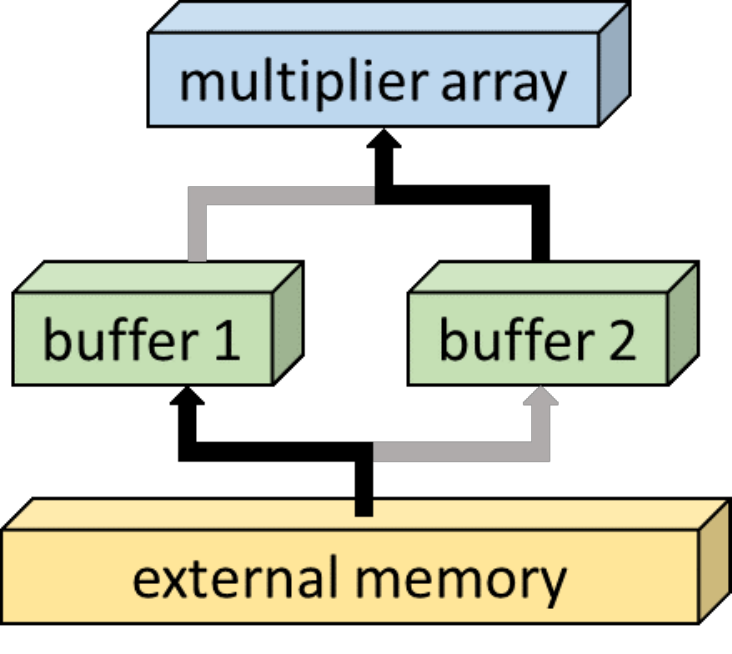}
		\label{fig:double_buffer_2}}
	\caption{Weight buffer in ping-pong structure}
	\label{fig:double_buffer}
\end{figure}

To have an efficient memory organization, one has to balance on-chip memory resources and external memory bandwidth. On-chip memory is limited on FPGA but supplies very high bandwidth. Contrarily, external memory
has the capability to store large amount of data but with the penalty of limited bandwidth. Therefore, in this proposed accelerator, we adapt the hierarchical memory methodology. Weight buffer loads the needed
parameters from external memory before each convolution starts. This, on one hand, reduces the latency caused by parameters loading, and on the other hand, avoids the latency brought the limited bandwidth of external memory. Besides, weight buffer is built as a ping-pong buffer (Fig.~\ref{fig:double_buffer}), which means that when weight buffer 1 outputs data for convolution, the weight buffer 2 loads the data from external memory for the next one and vice versa.
\par
Intermediate feature maps is another way chosen during system design to reduce processing time. Its size depends on the number of MME instantiated and the size of feature map.



\section{Results}\label{sec:system_impl}
The proposed accelerator architecture (Fig.~\ref{fig:sys_arch}) is demonstrated by implementing the MobileNetV2 network on the 
\textcolor{black}{Arria 10 SoC Development Kit} (10AS066N3F40E2SG), which contains 251680 ALMs, 2131 M20K, and 1687 DSP blocks. The design consideration will be described and then followed by implementation results with utilization.
\subsection{Implementation Consideration}
As mentioned in Section~\ref{sec:intro}, lower numerical precision is sufficient for CNN. So 16-bit quantization strategy is chosen because it is widely selected by previous works
\cite{chen2014dadiannao}\cite{du2015shidiannao}\cite{ma2016scalable}\cite{li2016high}.
\par
Based on the description in Section~\ref{sec:system_design}, 4-MME array is decided to instantiate in this design after carefully balancing the resources usage and processing time. The weight buffer size is 36Kb as a ping-pong buffer. Since the update rate of weights when performing depthwise separable convolution is every $M\times M$ clock cycles. The size of intermediate feature map buffer is 24.5Mb.

\subsection{Implementation Results}
\begin{figure}[htbp]
	\centering
	\includegraphics[width=2.7in]{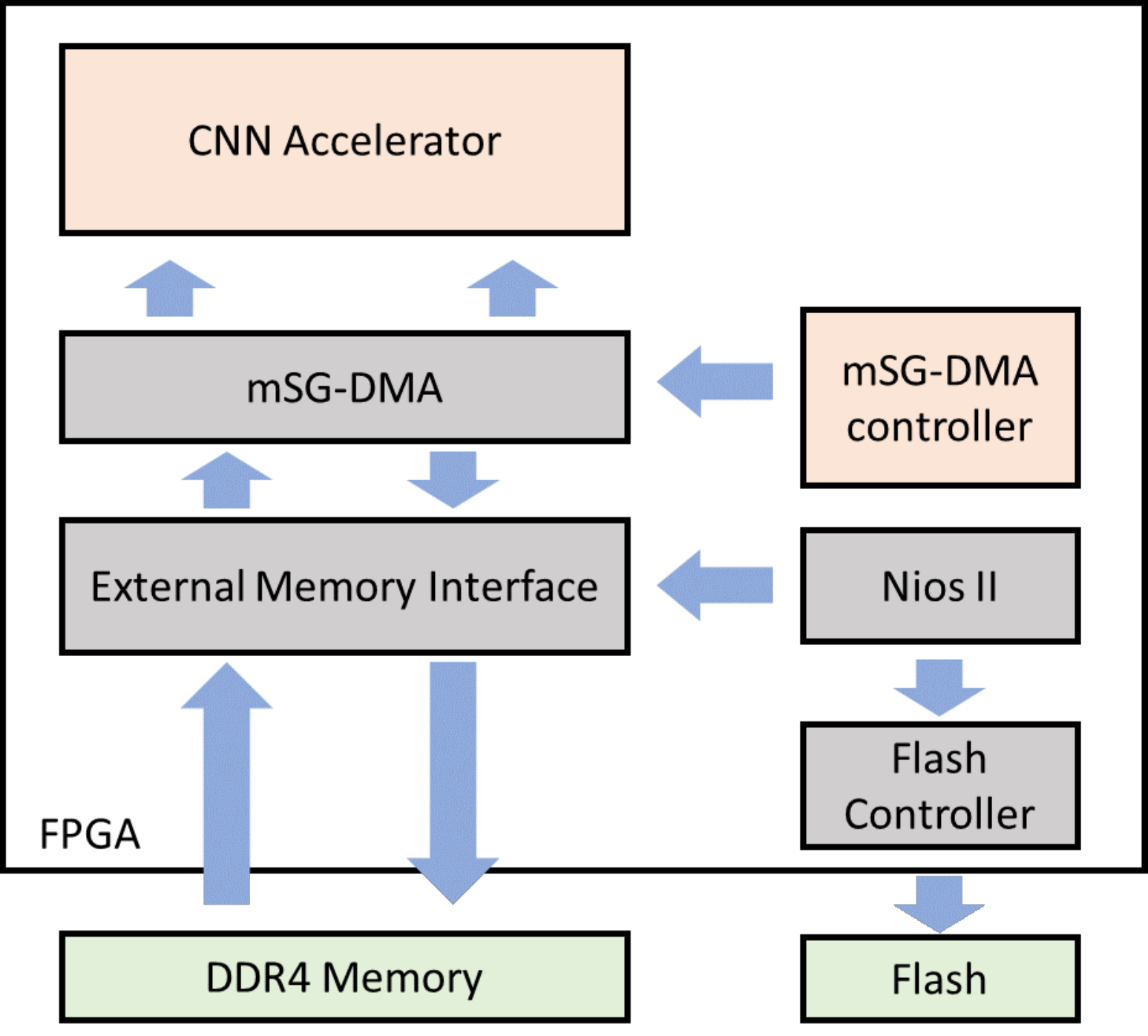}
	\caption{System architecture of the FPGA design}
	\label{fig:sys_arch}
\end{figure}
\textcolor{black}{Fig.~\ref{fig:sys_arch} presents the system architecture on Arria 10 SoC. Since HPS is not used in this design, only FPGA part is shown. The DDR4 memory is the one connected to the FPGA part.} The CNN accelerator runs at frequency 133MHz. Its adder tree limits this frequency. A Nios II softcore micro-processor is implemented for loading weights and input images from flash memory to 
DDR4 external memory.
An external memory interface IP combined with a Modular Scatter-Gather Direct Memory Access (mSG-DMA) IP are used to bridge the buffers in the CNN accelerator and the FPGA memory, whose maximum bandwidth is 8.5GB/s. This structure avoids the host's intervention during multiple 
transfers back and forth with DDR4 memory and makes non-continuous data movement more efficient.
The function of customized mSG-DMA controller makes it possible to drive mSG-DMA to read/write different sizes of data from/to specific addresses, in order to fit convolutions in various sizes.
\par
The implementation result is listed in Table~\ref{tab:impl_result}.

\begin{table}[htbp]
	\centering
	\caption{Resource Usage of MobileNetV2}
	\label{tab:impl_result}
	\begin{tabular}{|c|c|c|c|}
		\hline
		\textcolor{black}{Name} & \textcolor{black}{ALM} & \textcolor{black}{DSP} & \textcolor{black}{RAM} \\ 
		\hline
		\hline
		\textcolor{black}{MME} & \textcolor{black}{66127(26.3\%)} & \textcolor{black}{1278(75.9\%)} & \textcolor{black}{51(2.4\%)}\\ 
		\hline
		\textcolor{black}{Weight Buffer} & \textcolor{black}{9317(3.7\%)} & \textcolor{black}{0(0\%)} & \textcolor{black}{0(0\%)}\\ 
		\hline
		\textcolor{black}{Feature Map Buffer} & \textcolor{black}{1(0\%)} & \textcolor{black}{0(0\%)} & \textcolor{black}{1779(83.4\%)}\\ 
		\hline
		\textcolor{black}{Others} & \textcolor{black}{6308(2.5\%)} & \textcolor{black}{0(0\%)} & \textcolor{black}{14(0.6\%)}\\ 
		\hline
		\hline
		\textcolor{black}{Totally} & \textcolor{black}{81753(32.5\%)} & \textcolor{black}{1278(75.9\%)} & \textcolor{black}{1844(86.5\%)}\\
		\hline
	\end{tabular}
\end{table}
\textcolor{black}{Table~\ref{tab:cmp_result} provides a comparison between the solution proposed in this paper and other similar ones. Note that MobileNetV2 has more complex structure and higher accuracy on benchmarks.}

\begin{table}[htbp]
	\centering
	\caption{Comparison to other implementation}
	\label{tab:cmp_result}
	\begin{tabular}{|c|c|c|c|}
		\hline
		& \textcolor{black}{\cite{Jiang2018}} & \textcolor{black}{\cite{Zhao2018}} & \textcolor{black}{this paper} \\ 
		\hline
		\hline
		\textcolor{black}{Network} & \textcolor{black}{RR-MobileNet} & \textcolor{black}{MobileNetV1} & \textcolor{black}{MobileNetV2} \\
		\hline
		\textcolor{black}{Platform} & \textcolor{black}{Zynq UltraScale+} & \textcolor{black}{Stratix-V} & \textcolor{black}{Arria 10 SoC} \\
		\hline
		\textcolor{black}{Speed} & \textcolor{black}{127.4 fps} & \textcolor{black}{231.7 fps} & \textcolor{black}{266.2 fps} \\
		\hline
	\end{tabular}
\end{table}

\section{Conclusion}\label{sec:conclusion}
In this paper, a high-performance, scalable CNN accelerator is proposed. This
structure is optimized for depth separable convolution, which results in remarkably
less operations and parameters. This makes it possible to run the CNNs on portable devices. By choosing different number of MMEs and variable on-chip memories, this 
accelerator can be fit into a large or small FPGA. As an example, the latest MobileNetV2 is implemented on Arria 10 SoC 
FPGA, which achieves 266.6 fps and 170.6 GOPS.


%



\ifCLASSOPTIONcaptionsoff
  \newpage
\fi



%
\bibliographystyle{IEEEtran}




\end{document}